\documentclass[12pt]{article}

\usepackage{amsmath,amsfonts,amssymb}
\usepackage{graphicx}
\usepackage{psfrag}
\usepackage{enumerate}

\makeatletter \@addtoreset{equation}{section}

\makeatletter\renewcommand\section{\@startsection {section}{1}{\z@}%
                                   {-3.5ex \@plus -1ex \@minus -.2ex}%nn
                                   {2.3ex \@plus.2ex}%
                                   {\normalfont\large\bfseries}}
\renewcommand\subsection{\@startsection{subsection}{2}{\z@}%
                                     {-3.25ex\@plus -1ex \@minus -.2ex}%
                                     {1.5ex \@plus .2ex}%
                                     {\normalfont\bfseries}}

\newcommand{\be}{\begin{equation}}
\newcommand{\ee}{\end{equation}}
\newcommand{\beq}{\begin{eqnarray}}
\newcommand{\eeq}{\end{eqnarray}}

\def\[{\left [}
\def\]{\right ]}
\def\({\left (}
\def\){\right )}

\def\r2{\sqrt{2}}

\def\cc{cosmological constant}

\newcommand{\bbibitem}[1]{\bibitem{#1}\marginpar{#1}}

\newcommand{\figref}[1]{Fig. \ref{#1}}

\def\Label#1{\label{#1}%
  \smash{\hbox to0pt{\raise1ex\hbox{\tiny[#1]}\hss}}}
\def\noLabels{\let\Label=\label}
\def\nobbibitem{\let\bbibitem=\bibitem}

\begin{document}
%\noLabels % uncomment for final production
\nobbibitem % uncomment for final production

\begin{titlepage}

%\begin{flushright}%\vspace{-2cm}
%{\small
%UPR-1154-T  \\ %\vspace{-0.35cm}
%LBNL-60486 \\
%hep-th/0606118}%\\
%\end{flushright}
%\vspace{12 mm}

\vfil\
%vfil

\begin{center}

{\Large{\bf When Worlds Collide}}

\vspace{3mm}

Spencer Chang\footnote{e-mail:
chang@physics.nyu.edu}, Matthew Kleban\footnote{e-mail: mk161@nyu.edu} and Thomas S.
Levi\footnote{e-mail: tl34@nyu.edu}
\\

\vspace{8mm}

\bigskip\medskip
\centerline{\it Center for Cosmology and Particle Physics}
\smallskip\centerline{\it Department of Physics, New York University}
\smallskip\centerline{\it 4 Washington Place, New York, NY 10003.}

\vfil

\end{center}
\setcounter{footnote}{0}
%%%%%%%%%%%%%%%%%%%%%%%%%%%%%%%%%%%%%%%%%%%%%%%%%%%%%%%%%%%%%%%%%%%%%%%%%%%%%%%%%%%%%%%
\begin{abstract}
\noindent
We analyze the cosmological signatures visible to an observer in a Coleman-de Luccia bubble when another such bubble collides with it.  We use a gluing procedure to generalize the results of Freivogel, Horowitz, and Shenker to the case of a general cosmological constant in each bubble and study the resulting spacetimes.  The collision breaks the isotropy and homogeneity of the bubble universe and provides a cosmological ``axis of evil" which can affect the cosmic microwave background in several unique and potentially detectable ways.  Unlike more conventional perturbations to the inflationary initial state, these signatures can survive even relatively long periods of inflation. In addition, we find that for a given collision the observers in the bubble with smaller cosmological constant are safest from collisions with domain walls, possibly providing another anthropic selection principle for small positive vacuum energy.

\end{abstract}
%%%%%%%%%%%%%%%%%%%%%%%%%%%%%%%%%%%%%%%%%%%%%%%%%%%%%%%%%%%%%%%%%%%%%%%%%%%%%%%%%%%%%%%%%
\vspace{0.5in}

\end{titlepage}
\renewcommand{\baselinestretch}{1.05}  %Line spacing
%%%%%%%%%%%%%%%%%%%%%%%%%%%%%%%%%%%%%%%%%%%%%%%%%%%%%%%%%%%%%%%%%%%%%%%%%%%%%%%%
%%%%%%%%%%%%%%%%%%%%%%%%%%%%%%%%%%%%%%%%%%%%%%%%%%%%%%%%%%%%%%%%%%%%%%%%%%%%%%%%%%%%%%%%%%%
\tableofcontents

\newpage

\section{Introduction}\label{sec-intro}

Cosmology provides a tremendous opportunity to test theories of fundamental physics.  Recent progress \cite{discretum,kklt,lennylandscape} in string theory has led to a picture in which different patches of the universe correspond to different metastable vacua, each with its own local physics and value for the cosmological constant (CC).  These patches do not evolve entirely in isolation from each other.  Regions with the largest positive CC will inflate the fastest, and instantons connecting the different vacua will give rise to bubble nucleation.  Typically the nucleation rate is much slower than the rate of expansion, so that the bubbles do not percolate and the universe as a whole inflates eternally.  Even so, previously isolated bubbles will occasionally collide with each other, forming clusters.   

In this model our observable universe fits inside an isolated bubble, one which happens to have a very small CC and  low energy physics that corresponds to the Standard Model.  The bubble nucleation event in our past means that our universe evolved from very special initial conditions.  This leads to falsifiable predictions for cosmological observables, the sharpest of which is that the spatial curvature cannot be positive \cite{cdl,gott1,gott2,Garriga:1998he,lindeopen,Freivogel:2005vv, batra}. 

Unfortunately any theory of cosmological initial conditions suffers from the ``defect" that a sufficiently long period of inflation wipes away any memory of them.  In fact if inflation lasted only a few more efolds than are necessary to solve the standard cosmological horizon and flatness problems, it is unlikely that we will ever be able to make observations that significantly constrain the initial state.  It is not unreasonable to hope that inflation was short enough that this is not the case \cite{Freivogel:2005vv}, but nevertheless it is important to ask whether there are some other predictions of this model that may be visible even after significant inflation.

If indeed the landscape model for the universe is correct, the bubbles do not evolve in isolation from each other.  In fact, every bubble will eventually collide with an infinite number of others \cite{oldguth}.  The typical time between such collision events is model-dependent, and further seems to depend on a choice of otherwise equivalent observers \cite{ggv,Aguirre:2007an}.  Therefore it is difficult to make statements about how many bubble collisions to expect in the past lightcone of a generic observer.  Much remains to be understood concerning this question (both in this case and in the more general framework of the landscape) but we will have no more to say about it here.

In this paper we will focus on the cosmological signatures of a collision between two bubbles immersed in a ``sea" of inflating false vacuum, as seen by an observer living in the expanding universe inside one of the bubbles.  The collision event breaks the isotropy and homogeneity of the space inside the observer's bubble, picking out a preferred direction.  Such an anisotropy is rather difficult to achieve in more conventional models, since long inflation wipes away any initial inhomogeneity, and in Lorentz-invariant theories it is difficult to find dynamical effects which break isotropy on large scales.  The difference here is that the anisotropy is present at the ``big bang" but is not a small perturbation, and therefore is not necessarily removed by inflation.  Ordinarily such a large deviation from Friedmann-Robertson-Walker isotropy would render the problem intractable, but in this case we are able to solve for the metric analytically.  

In order to analyze the problem we solve Einstein's equations analytically in the thin-wall approximation, using a matching technique \cite{Israel:1966rt,israelnull} to generalize the results of Freivogel, Horowitz, and Shenker \cite{ben} (FHS) to all values of the \cc~inside each bubble (see also \cite{Hawking:1982ga,boussofreivogel}).  We solve for the motion of the domain wall separating the vacua.  One of the more intriguing aspects of our results is the fact that, in a collision between two bubbles of positive \cc, at late times the domain wall always accelerates {\em away} from any observer in the bubble with the {\em smaller} value of the cosmological constant.  These walls would otherwise be extremely dangerous, as they generically bound regions with very different low-energy physics and move at ultra-relativistic velocity.  Coupled with the result of \cite{ben} that walls connecting flat to anti-de Sitter that satisfy the BPS bound also accelerate away from the flat bubble, our results indicate that bubbles with very small positive or zero \cc~may be the only type that are safe from crunches and collisions with domain walls.  Since every bubble collides eventually with an infinite number of other bubbles of every possible type, this may mean that bubbles with the smallest positive value of the \cc~are ultimately the only safe places in the multiverse.  We leave further investigation of the implications of this to future work.

\section{Bubble backgrounds} \label{sec-back}
In this section we will outline the various spacetimes that form the building blocks for our solutions. Following the work of FHS \cite{ben}, we find that if we work in the thin wall approximation (both for the domain wall and the shell of energy it radiates) the spacetime resulting from the collision of two bubbles full of arbitrary vacuum energy can be constructed by gluing, with thin tensionfull walls localized at the gluing surfaces. 

For the collisions we consider we will need to consider vacuum solutions to Einstein's equations with cosmological constant that possess a particular hyperbolic invariance.  A single Coleman-de Luccia bubble in four dimensions has an $SO(3,1)$ invariance (inherited from the $SO(4)$ symmetry of the Euclidean instanton \cite{cdl}). When another bubble is present a preferred direction is picked out, breaking the symmetry down to $SO(2,1)$.  One can think of this as the Lorentzian version of the SO(3) invariance of a pair of marked points in four-dimensional Euclidean space.

As a result of this invariance, locally we need only to find the most general solution to Einstein's equations (with \cc) that has an $SO(2,1)$ invariance.  The situation is closely analogous to the more familiar case of spherical symmetry, and the metrics turn out to have the structure of a warped product of a two-dimensional hyperboloid with another two dimensional spacetime, and (for a given \cc) are characterized by one constant.  The 2-hyperboloids can in general be either spacelike or timelike, but will always be spacelike in the regions of interest to us. In the next section we present each background in some detail and analyze its conformal structure.

\subsection{dS space}
The most general metric for a spacetime with positive cosmological constant and hyperbolic symmetry is%
\beq \label{dsmetric}%
ds^2= -{dt^2 \over g(t)}+g(t)\, dx^2 +t^2\, dH_2^2 ,%
\eeq%
where%
\beq%
g(t) = 1+{t^2\over \ell^2}-{t_0 \over t},%
\eeq%
and%
\beq%
dH^2_2 = d\rho^2+\sinh^2\rho\, d\varphi^2%
\eeq%
is the metric on the unit hyperboloid. We will need to consider both positive and negative values for $t_0$. When $t_0=0$ the space is pure dS with radius of curvature $\ell$, written in hyperbolic coordinates. 

Before studying the general case it will be instructive to obtain the spacetime conformal diagram for plain de Sitter space in this slightly non-standard slicing. The full de Sitter manifold is the surface $-X_0^2+\sum_{i=1}^4 X_1^2 =1$ in five dimensions, with metric $ds^2=-dX_0^2+\sum_i dX_i^2$ (for the moment we set $\ell=1$, we will restore it in section \ref{sec-col}). In our coordinates the embedding is given by%
\beq%
X_0 &=& t\, \cosh \rho ,  \nonumber \\
X_1 &=& \sqrt{1+t^2}\, \sin x, \quad X_2 = \sqrt{1+t^2}\, \cos x , \nonumber \\
X_3 &=& t\, \sinh\rho\, \sin \varphi, \quad  X_4 = t\, \sinh\rho\, \cos \varphi.
\eeq%
To cover the upper part of the hyperboloid once we should take $x \sim x+ 2\pi$,  $0<t, \rho<\infty$, and $\phi \simeq \phi + 2 \pi$. To obtain the full manifold we need to take some of the coordinates off into the complex plane (otherwise points such as $X_0=0$, $X_2, X_3 \neq 0$ cannot be reached). As usual, this complexification happens when we cross the event horizons.  To reach the regions $X_3^2 + X_4^2 > X_0^2$ we continue both $\rho$ and $t$:%
\beq \label{dscomplex}%
\rho \to \rho + i\pi/2 \ \textrm{or} \ i 3\pi/2, \quad t \to - i \tau ,%
\eeq%
where $0\leq \tau \leq 1$, and in the region $X_0 <0,  X_0^2 > X_3^2 + X_4^2$ we take $\rho \to \rho+i \pi$ and leave $t$ real. A single cover again means $0<t<\infty$. 

\begin{figure}
\centering \hspace{0.2in}
\includegraphics[width=1.0\textwidth]{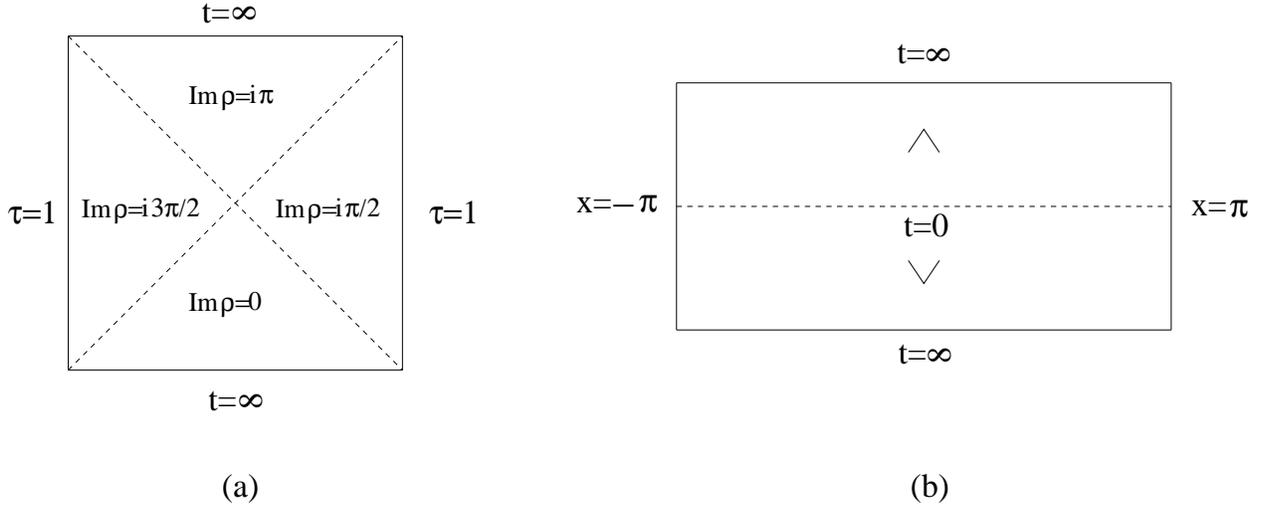} \caption{The Penrose diagram for hyperbolic dS space, (a) is the diagram suppressing $x, \varphi$ and (b) is the diagram suppressing the $H_2$. The wedges in each region are the Bousso wedges discussed in the text.
\label{dsm0}}
\end{figure}

We can draw more than one Penrose diagram depending on which coordinates we choose to plot. Since in our solutions nothing depends on the hyperboloid directions, ordinarily we want to suppress $\rho$ and $\varphi$ and plot the $t,x$ plane. However it is also instructive to consider suppressing $x, \varphi$ and plotting the $t, \rho$ plane. When we do so we can find a natural conformal time coordinate $T_\rho$ given by%
\beq%
T_\rho=\int_t ^\infty {dt' \over t' \sqrt{1+t'^2}}= - \ln \left({t\over 1+ \sqrt{1+t^2} }\right),%
\eeq%
and the metric becomes $ds^2 = t^2 (-dT_\rho^2+d\rho^2)$.  Because $-\infty < T < 0$ we can see that the section covered by real values of $T_\rho, \rho$ is the lower half of a diamond, and then using the transformations \eqref{dscomplex} we can complete the Penrose diagram in \figref{dsm0}(a).

The diagram we are more interested in is the $t,x$ plane. In this case the appropriate conformal time is%
\beq%
T=\int_0 ^t {dt' \over 1+t'^2} = \tan^{-1} t ,%
\eeq%
with a range $-\pi/2 \leq T \leq \pi/2$. The two-dimensional metric is $ds^2 = g(t) (-dT^2+dx^2)$ and is plotted in \figref{dsm0}(b).

A useful tool in constructing bubble collisions is the notion of ``Bousso wedges'' \cite{wedges}.
At any point in these spacetimes we can consider the four intersecting null rays in the plane perpendicular to the $H_2$.  Two of these four null rays will point along directions of decreasing radius of curvature of the hyperboloids, the other two will point along increasing directions. To draw the wedge, we draw a vertex ``$\cdot$'' at the point, and then extend lines from the vertex along the decreasing directions, e.g. if the radius of curvature decreases along both past directed rays we would draw the wedge ``$\wedge$''. For the $t_0=0$ dS space we have filled in these wedges for \figref{dsm0}(b) and we will display them for the other geometries as well.

\begin{figure} 
\centering \hspace{0.2in}
\includegraphics[width=0.3\textwidth]{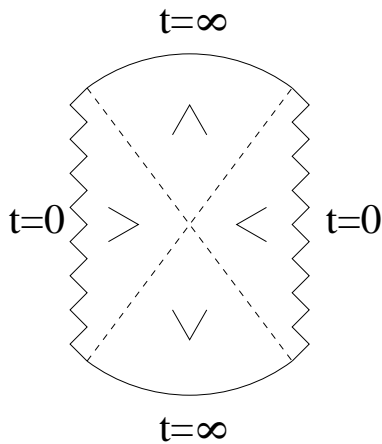} \caption{The Penrose diagram for the hyperbolic dS black hole suppressing the $H_2$. The wedges in each region are the Bousso wedges discussed in the text and the jagged lines are curvature singularities. \label{dsbh}}
\end{figure}

When $t_0>0$, we have a kind of hyperbolic Schwarzschild black hole in dS space, which has a horizon at $t=t_h>0$ (satisfying $g(t_h)=0$) and a timelike curvature singularity at $t=0$.\footnote{This spacetime is not stationary outside of the horizon due to the explicit time dependence, thereby evading Hawking's theorem that four dimensional black hole horizons must be spherical.} The Penrose diagram can be constructed using techniques similar to the ones we used for the $t_0=0$ case and is displayed in \figref{dsbh}.\footnote{For an explanation of why spacelike infinity must be drawn with a curved boundary see \cite{mattgeo}.} When $t_0<0$ the spacetime has a naked singularity. These spacetimes either begin with a Big Bang singularity or end in a crunch. We cannot make consistent collision scenarios using $t_0<0$ if we assume that the initial bubbles are expanding, and hence do not consider these spacetimes further here. We sketch the Penrose diagram in \figref{negmass}(a).
 
\subsection{Flat space}
\begin{figure} 
\centering \hspace{0.2in}
\includegraphics[width=0.7\textwidth]{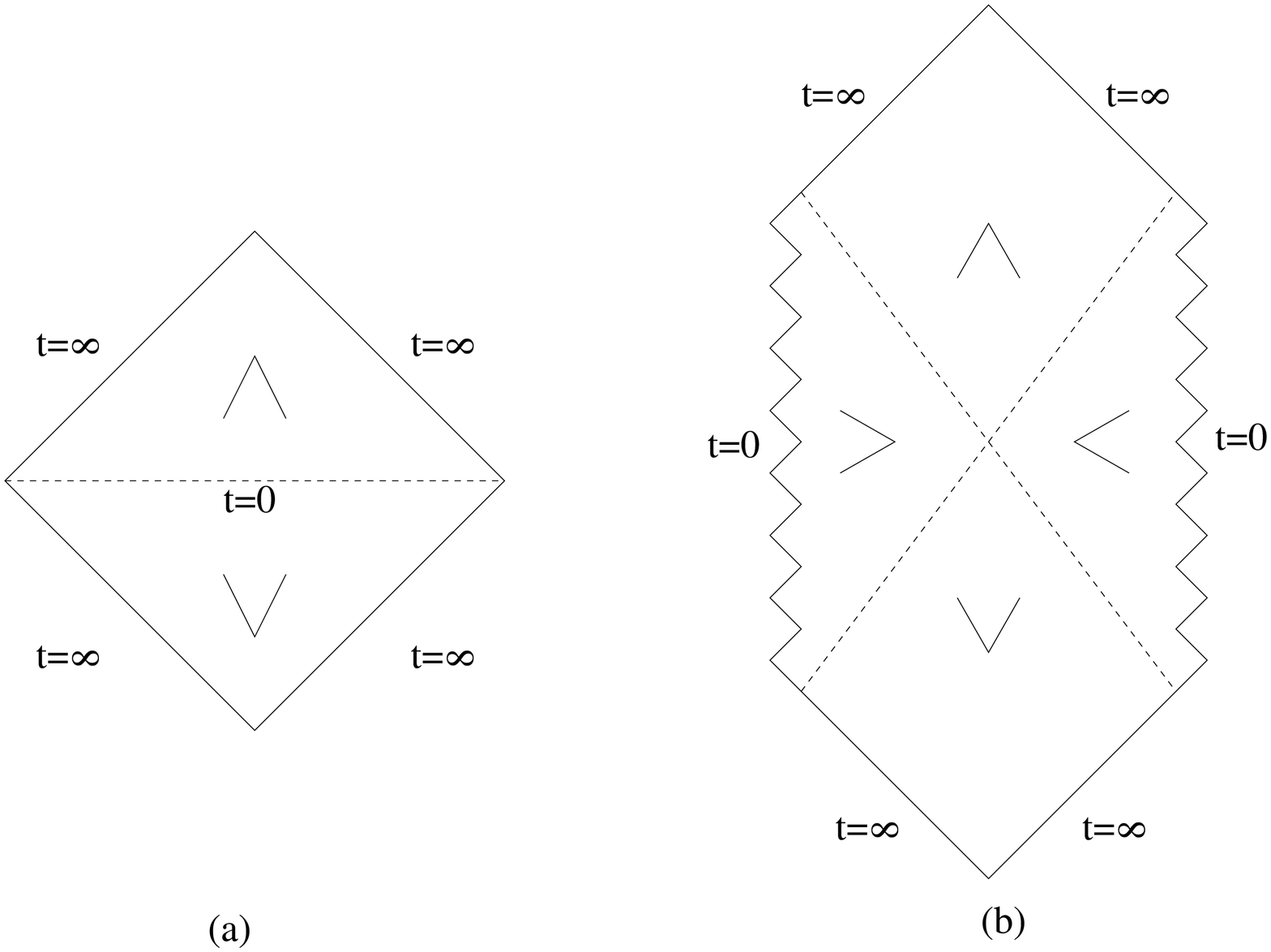} \caption{The Penrose diagram for hyperbolic flat space, (a) is the diagram for hyperbolic Minkowski space ($t_0=0$), (b) is the diagram for the hyperbolic Minkowski ``black hole'' ($t_0>0$). The wedges in each region are the Bousso wedges discussed in the text and the jagged lines are curvature singularities. \label{mink}}
\end{figure}

The unique solution for a hyperbolic spacetime with vanishing cosmological constant is given by%
\beq
\label{flatmetric}
ds^2 &=& -{dt^2 \over h(t)} + h(t)\, dx^2+t^2\, dH_2^2 , \\
h(t) &=& 1-{t_0 \over t} .%
\eeq%
When $t_0=0$ this metric gives flat space in Milne coordinates. The coordinate $t$ has range $0\leq t<\infty$, while $-\infty \leq x < \infty$. When $t_0>0$, the spacetime has a horizon at $t=t_0$ and a timelike curvature singularity at $t=0$. Using techniques similar to the ones we used for dS space, we can construct the Penrose diagrams in \figref{mink}(a) and (b). While one might worry about the presence of timelike singularities, in the next section we will see that for our collision scenarios we will only ever need to make use of the regular parts of the spacetime. When $t_0<0$ the spacetime has a naked, spacelike singularity at $t=0$ and we again find that we cannot find a consistent collision if the initial bubbles are expanding. We sketch the Penrose diagram in \figref{negmass}(b).

\subsection{AdS space}
\begin{figure} 
\centering \hspace{0.2in}
\includegraphics[width=0.3\textwidth]{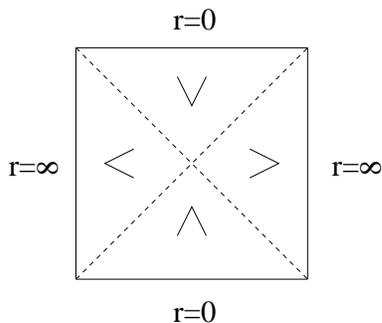} \caption{The Penrose diagram for AdS in the hyperbolic slicing. The wedges in each region are the Bousso wedges discussed in the text. \label{ads}}
\end{figure}

For hyperbolic space with a negative cosmological constant, the solution is%
\beq%
ds^2 &=& -f(r)\, dt^2+{dr^2 \over f(r)}+r^2\, dH_2^2 ,\\
f(r) &=& {r^2 \over \ell ^2}-1-{2 G M \over r} ,%
\eeq%
with ranges $-\infty < t < \infty$ and $0 \leq r <\infty$. This spacetime was described in detail in \cite{ben}, we will briefly summarize here. When $M=0$, this metric describes AdS in hyperbolic coordinates with radius $\ell$ and $r=0$ is a coordinate singularity. The Penrose diagram is displayed in \figref{ads}. When $M\neq0$, $r=0$ is a curvature singularity.

\begin{figure} 
\centering \hspace{0.2in}
\includegraphics[width=0.7\textwidth]{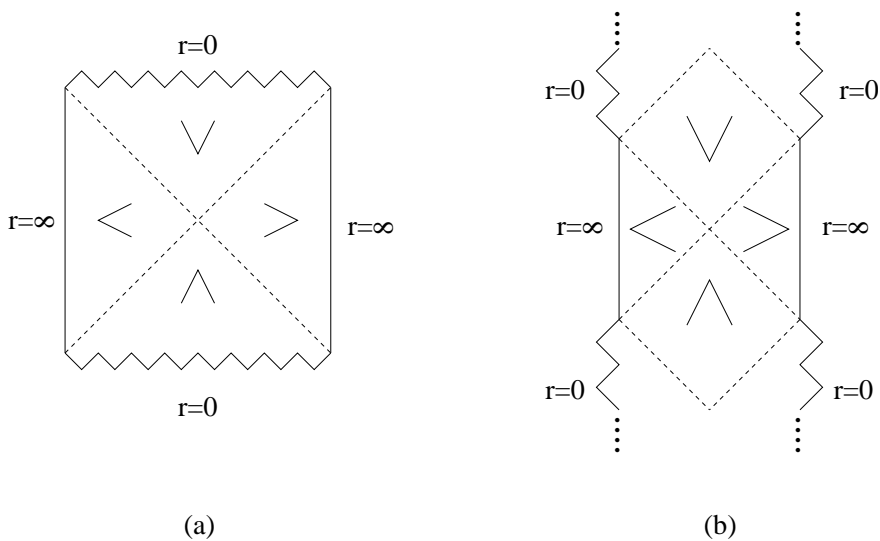} \caption{The Penrose diagram for the hyperbolic AdS black hole, (a) is the diagram for $M>0$ and (b) is the diagram for $-\ell / 3\sqrt{3}<GM<0$. The wedges in each region are the Bousso wedges discussed in the text and the jagged lines are curvature singularities. \label{adsbh}}
\end{figure}

\begin{figure} 
\centering \hspace{0.2in}
\includegraphics[width=0.7\textwidth]{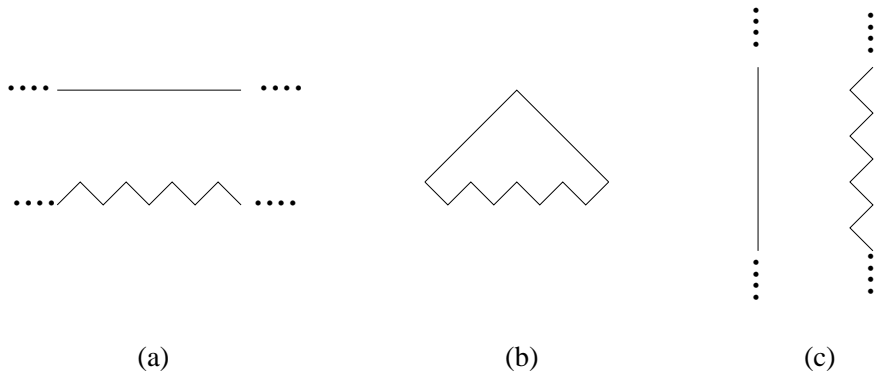} \caption{The Penrose diagrams for the nakedly singular spacetimes, (a) is the diagram for dS space with $t_0<0$, (b) is the diagram for flat space with $t_0<0$ and (c) is AdS space when $GM<-\ell / 3\sqrt{3}$. There are also the time reversed spacetimes.
\label{negmass}}
\end{figure}

When $M$ is such that $GM > -\ell / 3\sqrt{3}$ there is an infinite area hyperberbolic horizon at $f=0$. Hawking's theorem does not apply in this case because the energy density can be negative. When $M>0$ the singularities are spacelike and the Penrose diagram is shown in \figref{adsbh}(a). When $-\ell / 3\sqrt{3}<GM<0$ the singularities are timelike and there are two horizons, an inner Cauchy horizon as well as an outer event horizon located at the two zeros of $f$. The Penrose diagram is shown in \figref{adsbh}(b). Once again, we will see that we only need the regions of this spacetime that do not include the timelike singularity. When $GM<-\ell / 3\sqrt{3}$ the spacetime has naked singularities and we find that for expanding bubbles there is no consistent solution that uses this spacetime. We sketch the Penrose diagram in \figref{negmass}(c).

\section{Collisions} \label{sec-col}

In this section we will follow the notation and methodology of FHS \cite{ben} to construct consistent bubble collisions; readers who are familiar with that paper can skip ahead to section \ref{new}.  We briefly summarize the assumptions FHS took in their paper, which we also adopt:  

\begin{itemize}
\item{The thin wall limit is assumed for all bubbles.}
\item{The radiation radiated away into both bubbles by the collision is assumed to be null and confined to a thin shell.  It should be noted that numerical simulations in a similar context show that the radiation is emitted in a more complex pattern than this \cite{Hawking:1982ga,jose}; however we expect that a thin shell is a good approximation if the time scale of the collision is small compared to the timescale of the the bubble expansions and domain wall motion.}
\item{The collision forms a domain wall separating the two vacua, which we assume is thin.  The dominant source of energy on the domain wall is its relativistic tension, which is assumed to be positive.}
\item{The nucleated bubbles are initially expanding.}
\item{The Null Energy Condition is assumed.}
\end{itemize}

Given these assumptions, the collisions can be described by gluing together sections of the spacetimes presented in section \ref{sec-back} across the boundaries set by the null radiation shocks and the domain wall.  At each gluing surface Einstein's equations reduce to a junction condition which relates the discontinuity in the extrinsic curvature of the surface to the energy density.

\begin{figure} 
\centering \hspace{0.2in}
\includegraphics[width=0.4\textwidth]{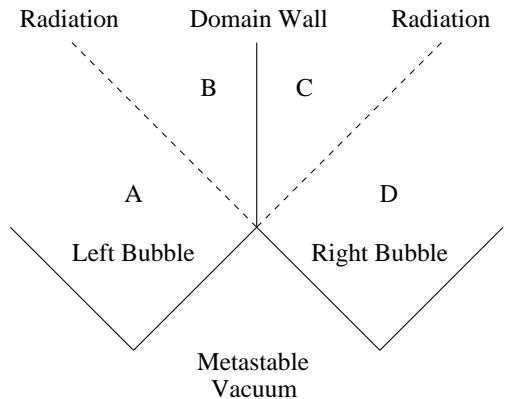} \caption{Schematic of a typical bubble collision event.  The dashed lines represent radiation emitted into the two bubbles by the collision.  After the collision a domain wall forms that separates the two bubbles. For future reference, we have labeled the different regions A, B, C, and D. The left (right) bubble is comprised of sections A and B (C and D).  \label{bubblecol}}
\end{figure}

To better vizualize the procedure, see \figref{bubblecol} for a schematic of a typical collision, with the radiation shocks and domain wall labeled (we have also labeled the different regions A, B, C, D for future reference.)  Solving the junction conditions across the radiation null lines will set the values of $M, t_0$ of the bubble geometry near the domain walls (regions B and C), and the junction condition at the domain wall (between B and C) determines the dynamics of the domain wall motion. 

The $SO(3,1)$ invariance of the individual Coleman-de Luccia bubbles is broken to $SO(2,1)$ by the axis connecting the nucleation points of the two bubbles.  We can use the broken boost generator of the $SO(3,1)$ to transform to a frame (the ``center of mass'' frame) in which the nucleation points are on a $t =$ constant slice.  A given observer comoving with the open FRW coordinates inside one of the bubbles will not be at rest in this frame, with a boost $\beta$.        

The fact that Coleman-de Luccia bubbles, when nucleated, are initially expanding determines what bubble metrics are allowed in regions A and D.  Regions A and D are parts of the bubble before the collision has occured and thus must be sections of the unperturbed flat, dS, or AdS solutions.  Therefore we should consider metrics for which $M, t_0 = 0$.  Since the bubble is expanding, the region of bubble nucleation must occur in a part of the spacetime where all future directed timelike geodesics have expanding $H_2$ radii.  

For flat or dS bubbles this confines the bubble nucleation to occur in the upper half plane of \figref{dsm0}(b) and \figref{mink}(a) with Bousso wedge $\wedge$.  (Note that this also determines that the bubble collision takes place in this region.)  For AdS bubbles, as seen from \figref{ads}, the nucleation occurs in the lower triangle with Bousso wedge $\wedge$, but the collision point can now occur in two regions.  If the AdS bubble is the bubble on the left, the expansion of its bubble wall to the right either stays in the region with Bousso wedge $\wedge$ or goes into the region outside of the horizon with Bousso wedge $>$.  Thus, for AdS bubbles there are two regions where the collision can occur.          

\subsection{Matching across the radiation shells}

We can match across the null shells and the domain wall as follows \cite{Blau:1986cw,Moss:1994iq,ben,israelnull}.  The metric is continuous across the junction for the transverse dimensions, which sets equal the radii of curvature of the $H_2$ hyperboloids, so $t$ or $r$ is continuous across the junctions.  Then the energy momentum at the junction determines the discontinuity in the extrinsic curvature.  

For the case of the null radiation shells in our thin shell approximation, the energy momentum tensor is
\beq%
T^{\mu\nu}= \sigma\, l^\mu l^\nu \delta ({\rm shell}) ,%
\eeq %     
where $l^\mu$ is a generator for the null line.  
The junction conditions impose the condition across the shell from the unperturbed metric (below) to one with nonzero $M, t_0$ (above), 
\beq%
\Delta k \equiv \left(h^{ab} k_{ab}\right)_{below} -\left(h^{ab} k_{ab}\right)_{above} = 8\pi G \sigma , %
\eeq%
where the normal vector $n^\mu$ that defines the extrinsic curvature $k_{ab}$ is future directed, normalized as $n\cdot l =-1$, and $h_{ab}$ is the spatial metric of the $H_2$ coordinates (including the $t^2$ or $r^2$ prefactor).  For flat and dS bubbles, there is only one case to consider since the collision occurs in a fixed region of the coordinate space.  For the flat case, we can show that 
\beq%
l^\mu = a(1,1/h,0,0) ,\quad \quad n^\mu=\frac{1}{2a}(h,-1,0,0) ,%
\eeq%
with $n$ future directed for positive constant $a$.  
This gives $k=h^{ab} k_{ab} = h/(a t)$ and the matching condition determines 
\beq%
\label{flatmatch}
t_0 = 8\pi G\, a\, t^2 \sigma(t) .%
\eeq%
The value of $a$ is somewhat arbitrary. Suppose at the time of collision an observer following the final domain wall has four-velocity $u^\mu$, and the hyperboloids have radius $R$. Then we can fix $a$ by requiring $u \cdot l=-1$, so that $\sigma(R)$ gives the energy density emitted into the flat bubble as seen by this observer. With this normalization, we find that $\sigma \sim 1/t^2$ as expected from energy conservation \cite{ben}. Similarly for the dS case, we get the correct result by simply replacing $h\to g$ giving the same equation as before
\beq%
\label{dSmatch}
t_0=8\pi G\, a\, t^2 \sigma(t) .%
\eeq%    

As we mentioned above, there are two cases in the AdS bubble depending on if the collision occurs inside or outside the horizon.  If it occurs outside of the horizon the tangent and normal vectors are
\beq
l^\mu = b(1/f,-1, 0, 0), \quad \quad n^\mu = \frac{1}{2b} (1,f,0,0) .%
\eeq    
Again, if $b$ is a positive constant, $n$ is future directed as required for our definition of the extrinsic curvature.  Calculating the extrinsic curvature, we find $k= f/(b r)$ and the junction condition gives
\beq
M = 4\pi \, b\, r^2 \sigma(r) .%
\eeq
We can choose the constant $b$ in a manner similar to that used for $a$, giving us $\sigma \sim 1/r^2$ which agrees with energy conservation. If the collision occurs inside the horizon, we have to take $n^\mu = -\frac{1}{2b}(1,f,0,0)$ to keep it future directed with $b>0$.  Thus,
\beq
M = -4\pi\, b\, r^2 \sigma(r) ,%
\eeq
so that a collision initiating inside the horizon creates a negative mass AdS black hole. While $M$ can be negative, we must have $GM>-\ell /3\sqrt{3}$ (otherwise the spacetime is nakedly singular), so the collision must satisfy $4\pi G b r^2 \sigma < \ell/(3\sqrt{3})$. 

\subsection{Matching across the domain wall}

To match across the domain wall we require that the induced metrics on both sides match up to a jump in the extrinsic curvature:%
\beq%
\Delta k^i_j = \left(k^i_j\right)_{left}-\left(k^i_j\right)_{right} = -8 \pi G( S^i_j - {1 \over 2} \delta^i_j S) ,%
\eeq % 
where $S_{ij}$ is the stress-energy tensor on the wall and $ij$ take values only over the three dimensions tangent to the wall. The normal vector for the extrinsic curvature is defined to point towards the right bubble. We assume that $S_{ij}$ takes the form of a perfect fluid and is dominated by a cosmological constant. This implies that $\rho+p=0$ and $\rho$ is constant. The junction condition becomes%
\beq%
\Delta k^i_j = 4\pi G \rho \, \delta^i_j \equiv \kappa \, \delta^i_j .%
\eeq % 
We can write the induced metric on the domain wall using proper time along the wall. On either side the induced metric takes the form
\beq%
ds^2_{\rm domain wall} = -d\tau^2 + R(\tau)^2 dH_2^2 ,%
\label{domainwallmetric}%
\eeq%
where $R(\tau) = r(\tau),\, t(\tau)$.  

The behavior of Bousso wedges across the junctions is another tool that FHS use to determine consistent bubble collisions.  
First of all, since the radii of the $H_2$'s are continuous across the null radiation lines, the Bousso wedges have to agree in the direction of the radiation line.  Furthermore, using the Null Energy Condition and Raychauduri's equation \cite{wedges}, one can show that along radially directed null lines where the $H_2$ is decreasing it must continue shrinking to zero size.  This determines that Bousso lines that ``cross'' a junction will continue in the same direction on the other side of the junction.\footnote{Since some of the spacetimes have a singularity when the $H_2$ shrinks to zero, we would encounter a problem if the junction was at such a singularity.  Fortunately, none of the consistent bubble collisions will have such junctions.}

Finally, it is useful to show that the junction condition across the domain wall can be analyzed as the classical motion of a particle in an effective potential.  The general junction condition imposes a constraint of the form 
\beq%
\eta_l \sqrt{\dot{R}^2+ j_l(\tau)} - \eta_r \sqrt{\dot{R}^2+ j_r(\tau)}= \kappa R ,%
\label{eq:generaljunction}
\eeq %    
where the subscripts $l,r$ denote the left, right bubble and for the different metrics $j(\tau) = -h(t) = -g(t) = f(r)$.  The signs $\eta_l, \eta_r = \pm 1$ in general are uncorrelated.  
In the dS and flat case, the extrinsic curvature is proportional to $dx/d\tau$, so the sign denotes whether the domain wall is moving to the left or right in the bubble coordinates.  While for the AdS case, the extrinsic curvature is proportional to $dt/d\tau$; outside of the horizon this sign is fixed and does not have the interpretation of the direction of the domain wall motion.    

Squaring this equation twice and solving for $\dot{R}^2$, we find%
\beq%
\dot{R}^2 = - V_{eff}(R) = -j_r(R) +\frac{[j_l(R)-j_r(R)-\kappa^2 R^2]^2}{4\kappa^2 R^2} .%
\label{eq:potential}
\eeq %
Therefore, the solutions can be analyzed by the motion of a particle with zero total energy in this effective potential.  
However, these solutions are not always consistent with the original equation \eqref{eq:generaljunction}, since the allowed signs $\eta_{(l,r)}$ may be fixed.

\subsection{de Sitter on de Sitter bubble collisions}\label{new}
We begin by considering the case of two de Sitter bubbles colliding.  To set conventions we will put the de Sitter bubble with larger $\Lambda$ on the left and denote it $\Lambda_l = 3/\ell_l^2$,  the bubble on the right has $\Lambda_r = 3/\ell_r^2$.  Let's first consider collisions which occur outside of the horizons (areas with Bousso wedge $\wedge$) in \figref{dsbh}; we have the junction condition at the wall
\beq%
\pm \sqrt{\dot{R}^2-g_l} \mp \sqrt{\dot{R}^2 -g_r}  = \kappa R,%
\eeq %
where the upper (lower) signs are when the domain wall is moving to the right (left) in each bubble.  All timelike geodesics that the domain wall can follow in the region with Bousso wedge $\wedge$ have $t\equiv R(\tau)$ increasing, so we can expand the effective potential for large $R$ by using $g_{l/r} \approx R^2/\ell_{l,r}^2$:
\beq%
\dot{R}^2 \approx \lambda^2 R^2 \quad {\rm where} \quad \lambda^2 \equiv \frac{1}{\ell_r^2}+\frac{1}{4\kappa^2}\left(\kappa^2 + \frac{1}{\ell_l^2}-\frac{1}{\ell_r^2} \right)^2 .%
\label{speed}
\eeq %   
The exponentially growing solution $R \sim R_0\, e^{\lambda \tau}$ is the correct solution, since all timelike geodesics in the regions must reach infinite time.    In general, the domain wall only moves a finite distance in $x$.  To see this, notice that the proper time defined in \eqref{domainwallmetric}, determines
\beq%
\dot{x}_{l,r}(\tau)=\pm \sqrt{-\frac{1}{g_{l,r}}+\frac{\dot{t}(\tau)^2}{g_{l,r}^2}} \approx \pm \, \frac{\ell_{l,r}}{R(\tau)}\, \sqrt{\lambda^2 \,\ell_{l,r}^2-1} ,%
\eeq%
where we have taken the large $R$ limit in the last step.  Since $R$ is exponentially growing in $\tau$, the net movement in $x$ is finite:%  
\beq%
\Delta x_{l,r} \approx \pm\, \frac{\ell_{l,r}}{\lambda \; R_0}\, \sqrt{\lambda^2 \,\ell_{l,r}^2 -1} .%
\label{xdist}
\eeq%
We can also determine the direction the domain wall moves, which depends on $\kappa$.  Putting the exponential solution for $R$ back into the original junction condition, we get the equation%
\beq%
\pm \frac{1}{2\kappa} \left|\kappa^2-\left(\frac{1}{\ell_l^2}-\frac{1}{\ell_r^2} \right)\right|\mp \frac{1}{2\kappa} \left|\kappa^2+\left(\frac{1}{\ell_l^2}-\frac{1}{\ell_r^2} \right)\right|   = \kappa .%
\eeq %
The terms in parentheses are positive since we assumed $\Lambda_l >\Lambda_r$ and thus the magnitude of the second term is larger than the first.  Since the right hand side is positive, we have to take the second sign to be a plus sign.  This means that the domain wall is moving to the left in the right bubble in the center of mass frame, and thus {\em in a de Sitter-de Sitter collision the domain wall always moves away from the bubble with smaller $\Lambda$.}

As for the bubble with larger $\Lambda$, there are three cases depending on $\kappa$.
\begin{itemize}
\item{$\kappa^2 > \frac{1}{\ell_l^2}-\frac{1}{\ell_r^2}$: This determines the first sign to be the upper sign, so the domain wall moves to the right, away from the left bubble.}
\item{$\kappa^2 < \frac{1}{\ell_l^2}-\frac{1}{\ell_r^2}$: The first sign is the lower sign, so the domain wall moves to the left, towards the left bubble. However, since the wall moves only a finite distance in $x$ we do not expect the left bubble to be eaten up, although part of infinity will be cut off.}
\item{$\kappa^2 = \frac{1}{\ell_l^2}-\frac{1}{\ell_r^2}$: The first term is zero, so the domain wall remains stationary at a fixed $x$.}
\end{itemize}

Under our assumptions such collisions are the only relevant de Sitter bubble collisions.  As already mentioned, we are only considering bubbles that are expanding in time.  In \figref{dsm0}(b), this means that regions A and D have a Bousso wedge $\wedge$.  Since $t$ is continuous across the null shells, we know that in region B (C), the Bousso wedge must have a portion $\searrow (\swarrow)$. Following the region B $\searrow$ across the domain wall determines that region C's Bousso wedge is $\wedge$.  Applying the same argument from region C to B determines region B's wedge to be $\wedge$.  In \figref{dsbh}, $\wedge$ Bousso wedges are only contained in the upper region outside of the horizon and thus the only consistent collisions occur outside of the horizon.  Such collisions are illustrated as type $iii$ in \figref{adscol}.    

If we are in a dS bubble with a small \cc~and we are hit by another such bubble with a larger one, the domain wall moves away from our bubble.  However if we are hit by one with a smaller cosmological constant, the domain wall will move away, stand still, or move towards our bubble if $\kappa^2$ is greater than, equal, or less than $1/\ell_{ours}^2-1/\ell_{other}^2$.  In all cases, the domain wall moves a finite coordinate distance as given by \eqref{xdist}.  

\subsection{de Sitter on flat bubble collisions}
A collision of a de Sitter with a flat bubble is simply the limit of the de Sitter on de Sitter bubble collisions when we take $\Lambda_r \to 0$.  The arguments for the collision to occur outside of the horizons of the hyperbolic de Sitter and Schwarzchild spacetimes go through as before, so there is again only one relevant solution.  As before the domain wall moves away from the flat space bubble, this time accelerating away to $x=-\infty$. Inside the de Sitter bubble, we have
\begin{itemize}
\item{$\kappa > \frac{1}{\ell_l}$: The domain wall moves to the right, away from the left bubble.}
\item{$\kappa < \frac{1}{\ell_l}$: The domain wall moves to the left, towards the left bubble.}
\item{$\kappa = \frac{1}{\ell_l}$: The domain wall remains stationary at a fixed $x$ in the left bubble.}
\end{itemize}
In all cases, the domain wall only moves a finite coordinate distance inside the de Sitter bubble.  Using these results, we see that if our dS bubble is hit by a flat bubble then the domain wall will move away, stand still, or move towards our bubble if $\kappa$ is greater than, equal, or less than $1/\ell_{ours}$. 

\subsection{de Sitter/flat on AdS bubble collisions}
\begin{figure} 
\centering \hspace{0.2in}
\includegraphics[width=0.8\textwidth]{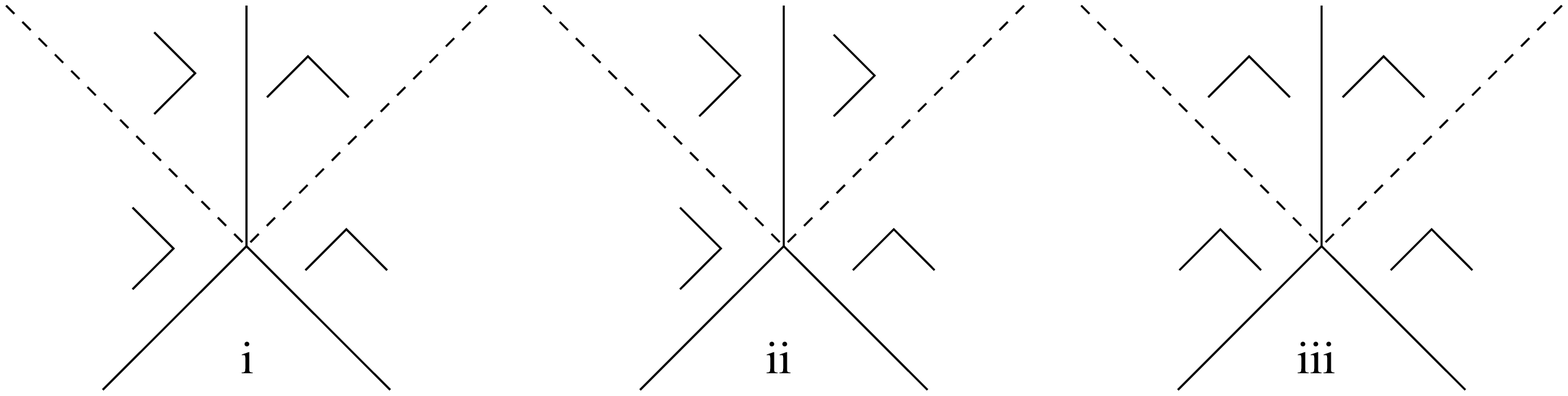} \caption{The three possible initial conditions of the collision of an AdS bubble on a flat/dS bubble.  The AdS (flat/dS) bubble is on the left (right) of each diagram.  Note that the Bousso wedges could change as the domain wall evolves into the future. \label{adscol}}
\end{figure}

A flat bubble colliding with an AdS bubble was already considered in FHS, but we will summarize the results to make this discussion self-contained.  We will also consider de Sitter bubbles colliding with Anti-de Sitter bubbles in this same section.  Let's first start by determining where the collision can occur by analyzing the Bousso wedges.  For this section, we will put the AdS bubble on the left and the flat (or dS) bubble on the right.  Unlike the previous cases, there are a couple of possibilities since on the AdS side the collision either occurs inside or outside of the horizon.  For the flat/dS side, the fact that the bubble is expanding already sets the initial Bousso wedge in region D to be $\wedge$.  Using the properties of the Bousso wedge across the different junctions, we see that this determines the Bousso wedge in A, B, and C to have the leg $\swarrow$.  Thus, the consistent wedges in these regions must be either $\wedge$ or $>$.  In fact, there are only three possibilities as shown in \figref{adscol}.  

Let's first describe the collisions of type i.  In this case, the regions B and C are outside of both the AdS horizon and dS/flat horizon respectively.  The junction condition at the domain wall is that $R(\tau)=r(\tau)=t(\tau)$ and %
\beq%
\sqrt{\dot{R}^2+f} \mp \sqrt{\dot{R}^2 -g} = \kappa R,%
\eeq%
where we assumed the collision with a dS bubble (for a flat bubble, replace $g$ with $h$).\footnote{The first term has only one sign choice because the extrinsic curvature for AdS is proportional to $\dot t$ which is positive for timelike domain walls outside of the horizon.}  The timelike geodesics in region C have $t$ monotonically increasing, so we can expand the effective potential in the large $R$ limit to get:%
\beq%
\dot{R}^2 \approx \lambda^2 R^2 \quad {\rm where} \quad \lambda^2 \equiv \frac{1}{\ell_{dS}^2}+\frac{1}{4\kappa^2}\left(\kappa^2 - \frac{1}{\ell_{AdS}^2}-\frac{1}{\ell_{dS}^2} \right)^2 .%
\eeq%
Plugging in the exponentially growing solution gives the junction condition%
\beq%
\frac{1}{2\kappa} \left|\kappa^2+\left(\frac{1}{\ell_{AdS}^2}+\frac{1}{\ell_{dS}^2} \right)\right| \mp \frac{1}{2\kappa} \left|\kappa^2-\left(\frac{1}{\ell_{AdS}^2}+\frac{1}{\ell_{dS}^2} \right)\right|  = \kappa ,%
\eeq%
where the upper (lower) sign for the second term indicates that the domain wall is moving to the right (left).  Inside the AdS bubble, the domain wall accelerates out to $r=+\infty$ in a finite time as can be seen by expressing $\dot{t}(\tau)$ in terms of $r(\tau)=R(\tau)$, hence the AdS bubble always ends in a crunch singularity.  The behavior inside the dS bubble is determined by the size of $\kappa$.     
\begin{itemize}
\item{$\kappa^2 > \frac{1}{\ell_{AdS}^2}+\frac{1}{\ell_{dS}^2}$: This determines the sign to be the lower sign, so the domain wall moves to the left, away from the right bubble.}
\item{$\kappa^2 < \frac{1}{\ell_{AdS}^2}+\frac{1}{\ell_{dS}^2}$: The sign is the upper sign, so the domain wall moves to the right, towards the right bubble. However, as before the bubble is not eaten up because the wall moves only a finite distance in $x$, although in this case too part of infinity will be cut off.}
\item{$\kappa^2 = \frac{1}{\ell_{AdS}^2}+\frac{1}{\ell_{dS}^2}$: The first term is zero, so the domain wall remains stationary at a fixed $x$ in the right bubble.}\end{itemize}
Again, in the dS bubble, the domain wall moves a finite coordinate distance of magnitude given in \eqref{xdist}.  We see that even though the AdS bubble crunches, the dS bubble will remain smooth. To determine the dynamics of an AdS bubble colliding with a flat bubble, take the limit where $\ell_{dS}\to \infty$.  In this case, the domain wall in the flat bubble can reach $x=\pm \infty$, eating up the asymptotic future of the bubble.  If $\kappa \geq \frac{1}{\ell_{AdS}}$, this never happens.  This also happens to be the BPS bound on the domain wall tension between two supersymmetric vacua (one flat, one AdS), so if the bound is satisfied for all such walls, flat bubbles are protected from being destroyed when colliding with crunching AdS bubbles \cite{ben}.  For our dS bubble, we see that after a collision with an AdS bubble the domain wall will move away, stand still, or move towards our bubble if $\kappa^2$ is greater than, equal, or less than $1/\ell_{ours}^2-1/\ell_{AdS}^2$.  The coordinate distance traveled is again given by \eqref{xdist}.        

Interestingly, it can be shown that both type ii and iii collisions ultimately evolve into the late time behavior of type i collisions.  For type ii collisions, as time evolves, region C will turn into a Bousso wedge of type $\wedge$.  To see this, note that for flat (dS) bubbles in \figref{mink}(b) (\ref{dsbh}), the timelike geodesics in the region with Bousso wedge $>$ either escape into the region with Bousso wedge $\wedge$ or they hit the timelike singularity.  We can see that the second possibility cannot happen.  Since $t=r$ at the domain wall, this means that $r$ approaches zero as well, but the only consistent timelike geodesics that go from the region with Bousso wedge $>$ in the AdS black hole space (see \figref{adsbh}(a)) towards zero $r$ reach a region with Bousso wedge $\vee$.  However, $\vee$ doesn't consistently match across the domain wall to the Bousso wedge $>$ in region C, thus such domain wall evolution does not occur.  Therefore, type ii collisions have the same late time behavior as type i collisions.  

Collisions of type iii are interesting because as discussed earlier, matching across the null shell between regions A and B, determines that $M$ is negative in region B.  These collisions also approach those of type i in the future.  This is because in region C, the timelike geodesics as seen in \figref{mink}(b) and \figref{dsbh} have $t$ monotonically increasing.  Since $t=r$ at the domain wall, this means that $r$ must go to infinity as time evolves.  Looking at \figref{adsbh}(b), the only timelike geodesics that go from the region with Bousso wedge $\wedge$ to infinity in $r$ go into the regions $<$ or $>$.  The wedge $<$ is incompatible with the wedge $\wedge$ in region C, thus in region B, the Bousso wedge ultimately turns into a $>$ as time evolves and thus has the same late time behavior as collision type i.           

\subsection{AdS on AdS bubble collisions}

AdS bubble collisions in general are not particularly relevant phenomenologically since we do not live in one.  However, the dynamics are of interest in and of themselves. The analysis of AdS bubble collisions follows the same general pattern as that presented above. However, there are more possibilities for the matchings than in the dS and flat cases and there can be more varied motion of the domain wall with respect to the two bubbles. We have analyzed all of these scenarios, but the end result is always that both bubbles end in some form of singularity, either as a result of hitting the spacelike singularity of the AdS black hole for $M>0$, the domain wall reaching asymptotic infinity in finite time and causing a big crunch singularity, or hitting a singularity that forms along the Cauchy horizon of the $M<0$ black hole. In the interest of brevity we will not present the details here.

\section{Signals}

Having written down explicit solutions describing the collision of two bubbles, we can use them to address the most interesting set of questions: what are the observable signatures of a bubble collision?  We will identify four possible signatures and discuss them qualitatively.  In a forthcoming paper we will perform a quantitative analysis of the effects in some specific models.   (See also \cite{Aguirre:2007an} for a discussion of some potential signatures of bubble collisions.)

In our approximation there are four quantities which determine the observable signatures: $t_0, R(t), t_c$, and $\beta$.  $t_0$ is the parameter in the metric \eqref{dsmetric} which is related to the gravitational effect of the shell of radiation and is determined by the amount of energy released at the collision.  $R(t)$ is the trajectory of the wall, determined by \eqref{eq:generaljunction}.  These two parameters are in principle computable given the potential for the scalar and the initial conditions.

The other two parameters are $t_c$, the value of the time coordinate at the collision point, and $\beta$, the boost of the observer with respect to the ``center of mass frame" (see \figref{signal}). These last two are initial conditions determined by the nucleation of the bubble.  As this is a quantum event, we could at best compute a probability distribution for these parameters.  We will not attempt to do so here.  

\subsection{Hitting the wall}

The first and most obvious signature visible to an observer inside one of the bubbles is a collision with the domain wall itself.  In the string theory landscape there are an enormous number of minima, which are generically endowed with string-scale vacuum energies and ``low energy" physics very different from ours.  As such, a collision with a domain wall separating our vacuum from another is likely to be extremely unhealthy.  Since the acceleration of the walls tends to be very high \eqref{speed}, their speeds are typically relativistic, and so there will likely be little warning if such an event is in our future.  

It is interesting to ask under what circumstances we should expect to undergo such a collision.  No matter what the trajectory of the domain wall is, some infinite class of otherwise cosmologically equivalent observers will have a $\beta$ such that they collide with the wall.  If the boost is such that the observer passes near the wall but doesn't quite collide with it, the effects discussed below will be greatly enhanced (but might remain nonlethal).  Therefore, in any collision event there are at least some observers who will observe a specific set of effects of essentially arbitrary strength.

\subsection{Radiation from the collision}

\begin{figure}
\centering \hspace{0.2in}
\includegraphics[width=.7\textwidth]{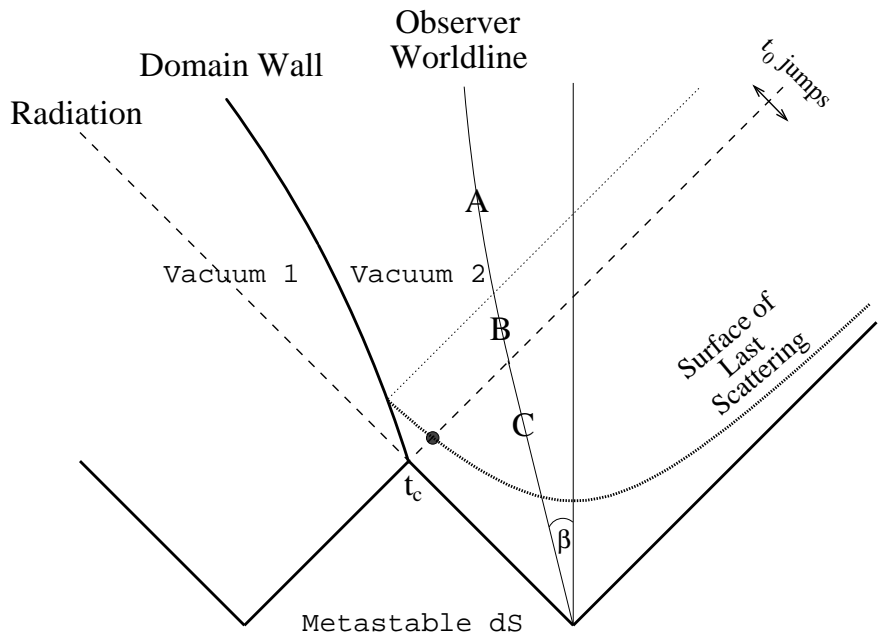} \caption{A conformal diagram showing the collision with the relevant signals regions labeled.  Observers in region A can see reflected photons off the domain wall (giving a Doppler-shifted disk in the CMB), those in region A or B see an angle-dependent perturbation spectrum in the CMB, and those in region C are unaffected by the collision. 
\label{signal}}
\end{figure}

When the two bubbles collide, radiation will be emitted inside a lightcone based along the collision hyperbola  (\figref{signal}).  In order to find the geometries analytically we have assumed that this radiation is confined to a thin shell.  In this case it affects the metric inside the shell because of its backreaction, and this jump in the metric will lead to potentially observable effects.  

Observing the radiation itself directly is difficult if it is concentrated in a thin outward-directed shell released prior to last scattering, as we have assumed in order to solve for the geometry.  However it is probable that the wall vibrates after the collision and continues to emit radiation which could potentially be observable.  One interesting fact is that gravitational and electromagnetic radiation is suppressed in the limit the walls are perfectly smooth, because the hyperbolic symmetry suffices to forbid any (for the same reason as for spherically symmetric configurations) \cite{Kosowsky:1991u}.   However perturbations will break this symmetry, and if the collision is sufficiently chaotic this may result in significant production of photons and gravity waves, along the hyperboloid as well as in the $x$ direction.  Such radiation would propagate inside the lightcone of the collision as well as along its edge, and gravity waves from this effect could form a potentially visible ring in the sky at late times, with an angular diameter anywhere between zero and half the sky.\footnote{We would like to thank S. Shenker and B. Freivogel for discussions on this point.}  Determining the type of radiation and its spatial distribution and spectrum will require a more detailed model.  Given a potential for the scalar, numerical simulations of the collisions should allow one to determine how much scalar radiation is emitted, and how thin a shell it radiates into. 

In order to estimate the size of the effects, the value of energy released in the shell can be solved for easily in a particularly symmetric situation.  Take the example of two bubbles of the same de Sitter vacuum in the center of mass frame.  Because the bubbles are nucleated on the same time-slice there is a reflection symmetry about the collision point.  When the bubbles collide, given our assumptions radiation is released symmetrically along the two lightcones, and the space above the radiation is a $t_0 \neq 0$ dS space with no domain wall.  In this case energy conservation determines the following condition at the collision point $t_c$ \cite{ben,Langlois:2001uq}
\beq
\cosh^{-1} \left(\frac{\dot{t}}{\sqrt{g_a}}\right)- \cosh^{-1} \left(\frac{\dot{t}}{\sqrt{g}}\right) = \frac{1}{2} \ln \frac{g_f}{g},
\eeq  
where 
\be
g_a = 1 + t^2/\ell_a^2, \quad g = 1 + t^2/\ell_b^2, \quad  g_f = 1-t_0/t + t^2/\ell_b^2,
\ee
describe respectively the metrics of the ambient metastable dS outside of the bubbles, the dS bubbles before radiation, and the final region after the radiation. The $\dot{t}$ terms represent the derivative of the bubble wall time coordinate with respect to its proper time, exactly the $\dot{R}$ terms discussed earlier for the domain wall.  Assuming that the collision time $t_c$ is large compared to $1/\ell_{(a,b)}$, the junction condition across the bubble wall gives 
\beq
\dot{t_c} \approx \lambda\, t_c \quad \quad \lambda^2 \equiv \frac{1}{\ell_a^2}+\frac{1}{4\kappa^2}\left(\frac{1}{\ell_a^2} - \frac{1}{\ell_b^2} -\kappa^2\right)^2,
\eeq    
which depends on $\kappa$, the tension of the bubble wall.  
Solving the energy conservation equation gives
\beq
\frac{g_f}{g}|_{t=t_c} &\approx & \exp\left[ 2 \cosh^{-1} (\lambda \ell_a) - 2\cosh^{-1} (\lambda \ell_b) \right]  \nonumber \\
 &=&\left(\frac{\lambda \ell_a+\sqrt{\lambda^2 \ell_a^2 -1}}{\lambda \ell_b+\sqrt{\lambda^2 \ell_b^2 -1}}\right)^2.
\eeq
There are some reasonable limits where the right hand side is small.  For example if we take $\lambda \ell_{(a,b)}$ large then the right hand side is $\ell_a^2/\ell_b^2$.  Similarly, since $\lambda \geq 1/\ell_{(a,b)}$, taking $\lambda$ near its smallest possible value $\ell_a$ gives a right hand side of order $\ell_a^2/\ell_b^2$.    
If there is a large ratio between the two cosmological constants, the right hand side can be small, which demonstrates that $t_0/t_c$ can be an effect of the same order as $t_c^2/\ell_b^2$.    

This shows that the perturbation to the metric inside the radiation shell is large close to the collision point.  However one might worry that a large number of efolds of inflation could undo this effect.  As we will see, this is not the case---or more precisely, for any given number of efolds of inflation $N$, if $t_c$ is sufficiently large the perturbation on the metric remains large even after inflation.

In order to see why this is the case, consider the diagram in \figref{signal}.  We have seen that the parameter $t_0$ is generically such that $t_0/t _c \sim t_c^2/l_b^2$.  Let us suppose the space inside the bubble continues to inflate, with Hubble constant $H_b = 1/l_b$ for $N$ efolds, and then reheats to a radiation dominated universe.  The relevant question is how large the perturbation in the metric due to the shell is at the reheating surface.  Reheating will occur along a line with a similar shape to the surface of last scattering in \figref{signal}. For the purposes of this discussion, we approximate the surface of last scattering as the reheating surface.

The value of $t$ at the point on the reheating surface where the null shell intersects it (the filled in circle on \figref{signal}).  In dS space, the constant $\tau$ slices of the $H_3$ metric describe a line in $(t,x)$
\beq
\cosh{\frac{\tau_{dS}}{l_b}} = \sqrt{1+\frac{t^2}{l_b^2}}\, \cos{\frac{x}{l_b}}.
\eeq
After N e-folds of inflation, we have $\sinh (\tau_{end} /l_b) \sim e^N$.  Solving for the null geodesic, the result (for large $t_c$) is that the time of crossing is given by $t_{cross} \sim (e^N/2)\, t_c$.  This means that 
\beq
\label{perturbed}
\frac{t_0}{t_{cross}} \sim e^{-N}\, \frac{t_0}{t_c} \sim e^{-N} \frac{t_c^2}{l_b^2}. 
\eeq
Since $t_c$ can be exponentially large, $t_0/t_{cross}$ can be very large.  To see how large the effect can be after inflation, we can match this inflationary de Sitter region onto an empty open bubble metric along the surfaces of constant $\tau_{flat}=\sqrt{t^2-z^2}$ and $\sqrt{1+(t^2/l_b^2)}\, \cos (x/l_b)$, where the matching occurs at\footnote{Notice that if $t_0/t_{cross} > 1$, after crossing into the post-inflationary metric, it appears behind the horizon of \figref{mink}(b).  The bubble solutions in this regime should require a full simulation to reliably determine their behavior.} $\tau_{flat}= l_b\, \sinh{(\tau_{dS}/l_b)}$.  Since the matching across the null radiation is the same for both dS and flat metrics \eqref{dSmatch} and \eqref{flatmatch}, $\sigma$  and $t$ are continuous at the transition between inflationary and post-inflationary regions.  In the post-inflationary region, $h(t) = 1-t_0/t$ after crossing the shell, thus $h(t)$ can have an order one shift when crossing the radiation shell.  Physical effects, like the redshifts of photons and geodesics of matter crossing the shell will receive corrections of order this difference.  We have also analyzed AdS on dS collisions and find that these physical effects can be large as well. Thus, in general, inflation does not wipe out potential signals.       

However there is a caveat to this statement.  In order to get an effect which can survive $N$ efolds of inflation, we needed to choose the collision time $t_c$ to be exponentially large in $N$.  However a glance at \figref{signal} makes it clear that the larger $t_c$ is, the later an observer at rest in the center of mass frame will come inside the lightcone of the collision and have a chance to observe the effects.  From our crude analysis (matching a de Sitter region onto a zero cosmological constant region) this constraint still allows an observer at rest to see a significant signal; using standard values for the cosmological parameters, we find $N<130$ for the effect to be $10^{-5}$ or larger.  However a more complete and detailed analysis will be necessary to determine this for certain.  Regardless, an observer sufficiently boosted towards the collision will always see a large effect.

When a true vacuum bubble nucleates it expands and accelerates.  If two bubbles nucleate far apart from each other, by the time they have collided the energy carried by the walls in the center of mass frame is high, and if the bubble we are in is inflating, it will have grown extremely large at that time.  As we have seen, a consequence of this is that the amount of energy emitted into the bubble can be very large.  This can be regarded as a large perturbation on the otherwise homogeneous initial conditions for the observer's bubble and presumably explains why even large amounts of inflation do not necessarily remove the effects.

\subsection{Mirror images}

A more interesting signature is possible when the domain wall is within reach of CMB photons.  For any collision, the surface of last scattering will intersect (or at least come very close to) the domain wall at some point.  If that point is within our past lightcone, some of the photons visible in the CMB today must either have reflected off the wall or been transmitted through it, as seen in \figref{signal} for the observer in region A.  
In general, if the wall separates regions with very different physics we expect it to have a reflection coefficient close to 1, at least for relatively low energy particles such as CMB photons \cite{Everett:1974bs}.  It is an interesting question to ask whether any particles will come through the wall.  

However, even if nothing is transmitted through the wall from the other side, reflected CMB photons have a very characteristic signature:  they will be doppler shifted by the motion of the wall.  This produces a circular disk in the CMB, inside of which the temperature is either red- or blue-shifted (depending on whether the wall is moving towards or away from us at the time of the reflection).  The doppler shift is a function of the radial distance from the center of the disk, and changes discontinuously at its edge which bounds the region where reflected photons arise from.  The edge will be smoothed out by the finite thickness of the surface of last scattering, but (if the doppler shift is significant and the disk not too small) this will create a very distinctive signature.

Due to the reflection, there will also be a correlation between pairs of concentric circles with radii smaller and larger than the radius of the disk, since these can correspond to nearby regions of the last scattering surface.  This effect is largest when the disk covers close to half the sky, but is probably not the most easily observed signature since the doppler shift is generically a larger effect.

These reflections will be present regardless of the amount of energy radiated by the collision, as long as the observer is sufficiently close to the wall and in region A  of \figref{signal}.

\subsection{Sachs-Wolfe}

The final effect we will consider is due to the gravitational backreaction of the shell of radiation.  As we have seen, the metric inside the shell is altered ($t_0 \neq 0$) and  accessible to any observer in region A or B in \figref{signal}.  One of the effects of this modification is that photons that originate inside the shell have a different integrated redshift from the surface of last scattering to us today than do those which originate outside.  In addition, photons that originate outside experience a redshift while passing through the shell which depends on their angle (relative to the direction of the collision), because they pass through the wall at different times when it has different energy density.  Hence the collision event determines a preferred axis of ``evil'' \cite{Land:2005ad, Hinshaw:2006ia} which points towards the collision and imprints itself on the CMB in a specific and calculable way.  

The entire sky will be divided into two disks by this effect: one in which the CMB photons last scattered outside the shell and then passed through it on their way to our instruments, and one in which the photons last scattered inside the shell, and did not pass through it afterwards.  In Figure \ref{signal} these disks will be centered on the $-x$ and $+x$ axes respectively.  
For the previously mentioned toy model of a sharp transition between an inflationary and post-inflationary open universe, we have calculated the differential redshift of photons that originated from outside the shell.  We find that this gives an effect of order $t_0/t_{observer}$, which (as mentioned earlier) even with significant inflation can be as large as $O(1)$ today and thus measurable.  Determining the full differential redshift on the sky requires solving for the surfaces of constant scalar field in the region inside of the shell, which we defer to future work.

\section{Discussion and future directions} \label{sec-con}

We have studied the dynamics and effects for general collisions between vacuum
bubbles. By gluing spacetimes together in the thin wall limit we have found
analytic solutions for the complete spacetimes and radiation. For collisions involving dS and flat
bubbles, the domain wall always accelerates away from the bubble with smaller
positive (or vanishing) cosmological constant at late times. For the other bubble,
the dynamics of the domain wall depend on its tension, which is
determined by microphysics. In particular, we find that AdS bubbles
always end in a crunch, but
the dS or flat bubbles they collide with can remain non-singular. We found
several distinct signals that can survive inflation and could be
detected by observables such as the CMB or other cosmological
measurements. The existence of an early phase of high-scale inflation
inside the observer's bubble will complicate this analysis, and it
would be interesting to work out the consequences for the domain wall
motion and signals.

There are many directions to go from here.  The task at hand is to perform a rigorous computation of the CMB power spectrum in the colliding bubble spacetime.  This will require solving the scalar field equations of motion in the region inside the null shell of radiation.  The spectrum will be characterized by the boost parameter $\beta$ of the observer with respect to the center of mass frame of the collision, the energy density released, and the time of the collision.  It will be very interesting to see whether this spectrum coincides with the ``axis of evil'' anisotropy which is present in the WMAP data \cite{Hinshaw:2006ia}.  It seems that, given the number of independent observables available in the data set, we should be able to fix the parameters and still test the model.

Many other cosmological observables will be affected by the collision.  Large scale structure formation, CMB polarization, gravitational lensing, and the 21cm spectrum are among the interesting possibilities.  If the collision created observable effects in the CMB temperature fluctuation spectrum, it is likely some of these other effects will be large enough to be observable as well.

A crucial question which we have not addressed is the likelihood such a collision is in our past lightcone and that the parameters take values which make it observable.  This is a confusing and subtle question on which we remain agnostic for the moment.  Given the potentially dramatic signatures, we prefer to compute the effects of the collision before attempting to address this problem.

A final point is our result that, in the collision of two de Sitter bubbles, observers in the bubble with the smaller value of the cosmological constant will see the domain wall move away from them (at least in the center of mass frame), thus increasing their chances of survival.  This may indicate the existence of an anthropic selection principle towards smaller values of the vacuum energy---particularly taken together with the results of \cite{ben}, which showed that flat bubbles are ``safe" from BPS bound-satisfying collisions with AdS bubbles, and our results for dS/AdS collisions, which again show that for fixed tension larger positive values of the \cc~are dangerous.  While it does not seem plausible that such an effect would predict as low a value for the vacuum energy as we observe, this effect may play a roll in the landscape measure.

\section*{Acknowledgements}
We would like to thank T. Banks, P. Batra, M. Blanton, R. Bousso, B. Freivogel, A. Gruzinov, L. Hui, A. MacFadyen, L. Randall, I. Sawicki, M. Strassler, S. Shenker, and L. Susskind for helpful discussions.  The work of M.K. is supported by NSF CAREER grant PHY-0645435.

\bibliographystyle{utphys}

\bibliography{bubble}

\end{document}